# Electron Localization Enhances Cation Diffusion in Reduced $ZrO_2$, $CeO_2$ and $BaTiO_3$

Yanhao Dong[1], Liang Qi[2], Ju Li[3], and I-Wei Chen[1*]

[1]Department of Materials Science and Engineering, University of Pennsylvania, Philadelphia, PA 19104, USA

[2]Department of Materials Science and Engineering, University of Michigan, Ann Arbor, MI 48109, USA

[3]Department of Nuclear Science and Engineering and Materials Science and Engineering, Massachusetts Institute of Technology, Cambridge, MA 02139, USA

**Abstract**

According to defect chemistry, the experimental observations of enhanced cation diffusion in a reducing atmosphere in zirconia, ceria and barium titanate are in support of an interstitial mechanism. Yet previous computational studies always found a much higher formation energy for cation interstitials than for cation vacancies, which would rule out the interstitial mechanism. The conundrum has been resolved via first-principles calculations comparing migration of reduced cations and oxidized ones, in cubic $ZrO_2$, $CeO_2$ and $BaTiO_3$. In nearly all cases, reduction alone lowers the migration barrier, and pronounced lowering results if cation's electrostatic energy at the saddle point decreases. The latter is most effectively realized when a Ti cation is allowed to migrate via an empty Ba site thus being fully screened all the way by neighboring anions. Since reduction creates oxygen vacancies as well, which are highly mobile, we also studied their effect on cation migration, and found it only marginally lowers the migration barrier. In several cases, however, a large synergistic effect between cation reduction and oxygen vacancy is revealed, causing an electron to localize in the saddle-point state at a much lower energy than normal, signaling that the saddle point is a negative-$U$ state in which the soft environment enables a large electron-phonon interaction that can over-compensate the on-site Coulomb repulsion. These general findings are expected to be applicable to defect-mediated ion migration in most transitional metal oxides.

*Corresponding Author Information

Tel: +1-215-898-5163; Fax: +1-215-573-2128

*E-mail address*: iweichen@seas.upenn.edu (I-Wei Chen)

*Postal address*: Department of Materials Science and Engineering, University of Pennsylvania, LRSM Building, Room 424, 3231 Walnut St., Philadelphia, PA 19104-6272

**7.1  Introduction**

Atomic migration to a neighboring vacancy strongly depends on the atomic size. A smaller atom migrates more easily than a larger one, because it can more easily pass through the crowded saddle point. Applying this argument to an ionic crystal, one would expect a reduced cation, which is larger in size, to experience more difficulty in migration. Meanwhile, the standard defect chemistry consideration also reaches the same conclusion on cation diffusion: Reduction increases the number of oxygen vacancies, so according to the Law of Mass Action—applied to the Schottky defect pairs, being vacancies of cation and anion—the number of cation vacancies decreases, leading to slower diffusion. However, there are experimental data that suggest the opposite: Cation diffusion is enhanced in reducing atmospheres as evidenced by enhanced grain growth in ceria[1-8], barium titanate[9,10] and yttria stabilized zirconia[2,11]. Resolving the above conundrum by first-principles calculations is the aim of this study.

Our study is first guided by the consideration on electrostatic energy, which is of paramount importance to the stability of ionic crystals. Therefore, we speculate that very likely it is also influential to migration barrier. A high-valence cation, while well screened in the ground state by anions, will probably become rather unstable during migration because the typical saddle-point state does not provide good screening. In this respect, a reduced cation is advantageous because its electrostatic energy is lower. Thus, our first task is to determine the migration barrier of a reduced cation by first-principles calculations.

From a practical viewpoint, the most interesting case for this study is when cation is the slowest diffusion species[12-15]. Obviously, this implies that other species, or rather, their vacancies, will migrate much faster. Therefore, if their presence can facilitate the migration of the slowest cation, then the latter can afford to wait for such vacancies to arrive before it migrates. Oxygen vacancies are abundant in $ZrO_2$, $CeO_2$ and $BaTiO_3$, and they are the fastest diffusing species. Therefore, our second task is to investigate the migration of Zr, Ce and Ti cations next to an oxygen vacancy. As it will become clear later in this study, Ba vacancies, which occupies the A-site sublattice of the perovskite structure as opposed to the B-site sublattice occupied by Ti, exert a most dramatic effect on Ti migration[16,17]. This effect will be studied, as well as the synergism between cation reduction and oxygen vacancy, to explore defect and ionization enhanced diffusion[18,19] and to understand the roles of charge screening and the saddle-point environment.

**7.2  Methodology**

To simplify matter, we considered cubic $ZrO_2$, $CeO_2$ and $BaTiO_3$, to avoid the complication of dopants and lattice distortions. (Typically, $Y_2O_3$ is added to $ZrO_2$ to provide oxygen vacancies and to stabilize the cubic structure, whereas $BaTiO_3$ is known to be ferroelectric below 120 °C forming several distorted structures. In our calculation, with and without defects, the cubic structure was found metastable, even though the size and shape of the supercell were allowed to relax when defects were introduced. Subsequently, during the migration calculation, the size and shape of the supercell were kept the same as those of the ground state.) We performed density functional theory (DFT) calculations using the projector augmented-wave (PAW)[20] method within the Perdew-Burke-Ernzerhof (PBE)[21] generalized gradient approximation (GGA) implemented in the Vienna *ab initio* simulation package (VASP)[22]. The PAW potentials include the following valence electrons: $5s^24d^2$ for Zr, $5s^25p^64f^15d^16s^2$ for Ce, $5s^25p^66s^2$ for Ba, $3s^23p^63d^24s^2$ for Ti and $2s^22p^4$ for O. We chose a plane-wave cutoff energy of 400 eV and sampled the Brillouin zone using the Monhorst-Pack scheme with a 3×3×3 *k*-point mesh. The DFT+*U* approach by Dudarev *et al.*[23] was used to describe the energy of localized *d* electrons of Zr/Ti and *f* electrons of Ce. Specifically, we chose the on-site Coulomb interaction parameter *U*, the on-site exchange interaction parameter *J*, and the effective Hubbard parameters $U_{eff}=U-J$ as follows: $U$=4 eV, $J$= 0 eV and $U_{eff}$=4 eV for Zr 4$d$ states[24-26], $U$=5 eV, $J$=0 eV and $U_{eff}$=5 eV for Ce 4$f$ states[27,28], $U$=5 eV, $J$=0.64 eV and $U_{eff}$=4.36 eV for Ti 3$d$ states[29-32]. These are the most commonly used values for the respective compounds in the literature, so our results may be directly compared with the literature results.

All calculations were performed under periodic boundary conditions. We used a 2×2×2 supercell containing 32 Zr or Ce and 64 O for cubic $ZrO_2$ and $CeO_2$, respectively, and a 3×3×3 supercell containing 27 Ba, 27 Ti and 81 O for cubic $BaTiO_3$. The migrating cation is situated next to a cation vacancy (in our case, a fully charged $V_M'''$, M being Zr, Ce or Ti, denoted as $V_M$ hereafter). Reduction, when considered, was implemented by providing an extra electron to the supercell. In some models, another surrounding oxygen vacancy ($V_O^{\cdot\cdot}$), or a A-site cation vacancy ($V_{Ba}''$) in perovskite, is also present. Therefore, it makes sense to preselect a target cation for migration, then to promote the localization of the extra electron around it. Toward this purpose, we outward-displaced the neighboring oxygen ions surrounding the target cation by 0.1-1.0 Å, then let the system relax to reach convergence (residue atomic forces less than 0.05 eV/Å). As will be shown later, the above process leads to obvious electron localizations in $ZrO_2$ and $CeO_2$ but not always in $BaTiO_3$ unless the 110-direction migration is considered.

To track cation migration, the climbing-image nudged-elastic-band (NEB) method[33] implemented in VASP was used with a fixed supercell size and shape. In cubic $ZrO_2$ and $BaTiO_3$, it determined the migration path and the barrier with the path defined by 7 intermediate states in addition to the initial and final configurations; in $CeO_2$, 3 intermediate states were specified. Convergence for NEB calculations was considered to be achieved when the residue atomic forces are less than 0.1 eV/Å.

DFT calculations always specify the Fermi level, which is used to determine electron state occupancy. However, to compare electron states in different structural states, such as ground state and saddle-point state, or states with and without some lattice ions, it is necessary to find a reference energy level that is relatively insensitive to the structural defects/distortions. All the structures studied here comprise of cation polyhedron enclosed by oxygen ions that are interconnected into a three-dimensional continuous network. Moreover, their valence bands are mainly made of O 2*p* orbitals. Therefore, we may regard the valence band manifold as representative of network's electronic states, and being a continuous network its overall electronic energy should be relatively insensitive to isolated structural defects/distortions. In this work, whenever needed we shall use the valence band maximum as the reference energy to compare electron energies between different structural states.

## 7.3 Cation Migration in $ZrO_2$ and $CeO_2$

To compare migration of $M^{4+}$ and $M^{3+}$, with or without the aid of oxygen vacancy $V_O^{\cdot\cdot}$, we follow the scheme in **Fig. 7.1a-b** (figure production uses software VESTA[34]). The calculated energy profiles during migration are shown in **Fig. 7.1c** and the key data are summarized in **Table 7.1**.

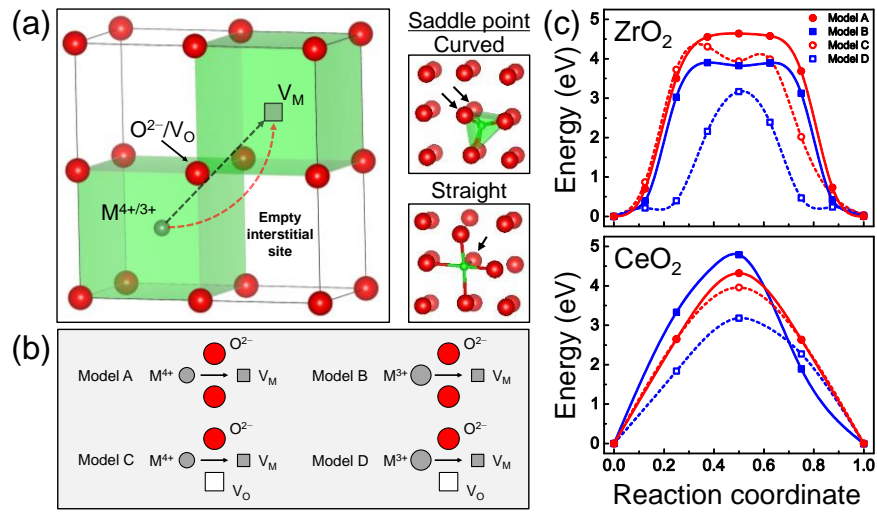

**Figure 7.1** (a) Schematics of 1×1×½ supercell with one cation vacancy $V_M$ for fluorite structure oxide $MO_2$ (M=Zr for zirconia and Ce for ceria). Cation migration takes place by exchanging the location of $M^{4+}/M^{3+}$ and $V_M$, with/without the assistance of an oxygen vacancy $V_O$. Two migration paths are found: straight path along black dashed line and curved path along red dashed line, their saddle point configurations shown on right panel. (b) Schematic migration models. **Model A**: $M^{4+}$ exchanges with $V_M$, with no $V_O$. **Model B**: $M^{3+}$ exchanges with $V_M$, with no $V_O$. **Model C**: $M^{4+}$ exchanges with $V_M$, with $V_O$. **Model D**: $M^{3+}$ exchanges with $V_M$, with $V_O$. (c) Cation migration energetics along 110 direction for **Model A-D** in cubic $ZrO_2$ and $CeO_2$.

**Table 7.1** Results of cation migration in cubic $ZrO_2$ and $CeO_2$ via vacancy mechanism. In our DFT+$U$ calculations of defect-free structures, reference Zr-O bond length is 2.24 Å, and Ce-O is 2.36 Å. With cation vacancy in the ground state, these bond lengths shorten. To calculate Ce's Bader charge, we exclude 8$e$ from $5s^25p^6$ inner-shell.

| Material | Model | Migration path | Migration barrier (eV) | Nearest cation-oxygen distance (Å) | | Bader charge on migrating cation ($e$) | |
|---|---|---|---|---|---|---|---|
| | | | | Ground state | Saddle point | Ground state | Saddle point |
| $ZrO_2$ | A | Curved | 4.64 | 2.15 | 1.97 | 0.62 | 0.80 |
| | B | Curved | 3.90 | 2.23 | 2.04 | 1.44 | 1.28 |
| | C | Straight | 4.31 | 2.15 | 2.05 | 0.59 | 0.68 |
| | D | Straight | 3.17 | 2.24 | 2.11 | 1.43 | 1.49 |
| $CeO_2$ | A | Curved | 4.32 | 2.27 | 2.03 | 1.63 | 1.74 |
| | B | Curved | 4.79 | 2.39 | 2.12 | 1.91 | 1.92 |
| | C | Straight | 3.96 | 2.25 | 2.06 | 1.65 | 1.67 |
| | D | Straight | 3.28 | 2.38 | 2.13 | 1.91 | 1.92 |

**Model A ($M^{4+}$ migration):** In this reference case, the migration path does not follow a straight line (the black dashed line in **Fig. 7.1a** from the starting $M^{4+}$ location to $V_M$). Instead, it veers into a cation-absent neighboring cell (the red dashed curve in **Fig. 7.1a**) to avoid the two oxygen ions in the way. The saddle point may be regarded as surrounded by six oxygens (**Fig. 7.1a**, upper inset); obviously, the two pointed by black arrows are closer than the remaining four. More broadly, the following general features listed in **Table 7.1** apply to all the models to be described later: (a) Zr is smaller than Ce, (b) Zr having considerably less Bader charge[35] is more ionic than Ce, and (c) the shortest M-O distance at the saddle point is considerably less than that at the ground state.

**Model B ($M^{3+}$ migration):** The migration path of $M^{3+}$ again veers into a cation-absent neighboring cell. We also verified the extra electron is indeed localized at the target cation and causes an increase of its Bader charge (see **Table 7.1**) in both the ground state and the saddle-point state compared to **Model A**. Consistently, a larger sized $M^{3+}$ compared to $M^{4+}$ is reflected in the longer M-O bond length. The localized electron lies just beneath the Fermi level in **Fig. 7.3a&d** and **Fig.**

**7.7a&d,** occupying a sharp, narrow projected density of state (DOS) of the migrating $M^{3+}$ in both the ground-state and the saddle-point state, falling between the valence band maximum (VBM) and the conduction band minimum (CBM) in **Model A**. Since it is clearly associated with the target cation (**Fig. 7.3b&e** and **Fig. 7.7b&e**), it is best regarded as an impurity state. When the ground state changes from $M^{4+}$ to $M^{3+}$, the increase in Bader charge is especially large for Zr, being 0.82*e*, compared to Ce's 0.28*e*. This is likely to result from the following well-known property of $Zr^{4+}$: According to Pauling's rule, $Zr^{4+}$ (but not the larger $Ce^{4+}$) is too small to be fully stable in the ground-state fluorite-structure environment of 8 oxygen neighbors, so it favors acquiring an electron forming $Zr^{3+}$, thus increasing the size and gaining stability. As the coordination number decreases to 6 at the saddle point, there is no more such need, so it sheds some localized electron as seen in **Fig. 7.3e** (DOS of the impurity state decreases from the ground state to the saddle-point state) and **Table 7.1** (Bader charge decreases from 1.44*e* to 1.28*e*). This does not apply to $Ce^{3+}$, but because of its larger size it encounters a higher migration barrier (4.79 eV compared to $Ce^{4+}$'s 4.32 eV) in passing through the crowded saddle point. In contrast, the migration barrier of $Zr^{3+}$ (3.90 eV) is actually significantly lower than that of $Zr^{4+}$ (4.64 eV). This gives the first indication that reducing the cation charge can facilitate cation migration.

**Model C ($M^{4+}$ migration aided by $V_O^{\cdot\cdot}$):** Upon removal of one intervening lattice oxygen, $M^{4+}$ now migrates along a straight line, with a somewhat lower migration barrier than that in **Model A**, by about 0.35 eV. Therefore, $V_O^{\cdot\cdot}$ apparently helps cation migration in both zirconia and ceria, which may be attributed to the size effect. The saddle point may be regarded as surrounded by five oxygens (**Fig. 7.1a**, lower inset); obviously, the one pointed by black arrow is closer than the remaining four.

**Model D ($M^{3+}$ migration aided by $V_O^{\cdot\cdot}$):** Like $M^{4+}$ in **Model C**, $M^{3+}$ also migrates along a straight line. Again, we confirmed the extra electron is localized at the migrating cation with the data in **Table 7.1**: (i) a larger Bader charge of the migrating $M^{3+}$ vs. $M^{4+}$ in **Model C** at both the ground state and the saddle-point state; and (ii) a longer M-O bond length of the migrating $M^{3+}$ compared to that of $M^{4+}$ in **Model C**, again at both states. **Table 1** also shows that, compared to **Model B**, the barrier is 0.73 eV lower in $ZrO_2$ and 1.51 eV lower in $CeO_2$, suggestive of a very significant size effect due to $V_O^{\cdot\cdot}$, which is especially important in $CeO_2$. But the benefit of reduction on lowering the barrier is equally significant: Compared to **Model C**, the barrier is lower by 1.14 eV in $ZrO_2$ and 0.68 eV in $CeO_2$, suggestive of a charge effect that is especially important in $ZrO_2$. This is supported by **Fig. 7.5b&e** and **Fig. 7.9b&e**, which show that during the migration the impurity state of the localized electron is pulled down toward the VBM—in the case of $Ce^{3+}$, it already merges into the valence band (O $2p$ orbitals) in the saddle-point state. This lowering in the impurity energy-level is about 1.5 eV in $Zr^{3+}$ and 1.2 eV in $Ce^{3+}$. Naturally, with such pronounced energy reduction in $ZrO_2$, there is no longer any electron shedding when $Zr^{3+}$ goes to the saddle point, so the Bader charge of $Zr^{3+}$ does not decrease in **Model D**, unlike in **Model B**. (Such significant lowering in the impurity energy-level from the ground state to the saddle-point state is not seen in **Model B**.)

A closer look finds the effects of $V_O^{\cdot\cdot}$ and reduction are synergistic. To see this, we first estimate their additive effect: The amounts of barrier lowering from **Model A→B** and from **Model A→C** add up to 1.05 eV for $ZrO_2$ and −0.11 eV for $CeO_2$. But the lowering of **Model A→D** is much larger: 1.47 eV for $ZrO_2$ and 0.94 eV for $CeO_2$. Therefore, the effect is not additive but synergistic. The synergistic effect is understood: Although the removal of one lattice oxygen creates a more open pathway, it also leaves the positive charge of the migrating cation much less screened, which raises the electrostatic energy. Therefore, by first receiving an extra electron and localizing it at the migrating cation, the electrostatic energy is lowered and the open space created by additional $V_O^{\cdot\cdot}$ may be utilized for migration. Because of the synergistic effect, the migration barrier of $Zr^{3+}$ with $V_O^{\cdot\cdot}$ is lowered to 3.17 eV compared to that of $Zr^{4+}$ without $V_O^{\cdot\cdot}$ at 4.64 eV, and corresponding values for $Ce^{4+}$ are 3.28 eV vs. 4.32 eV.

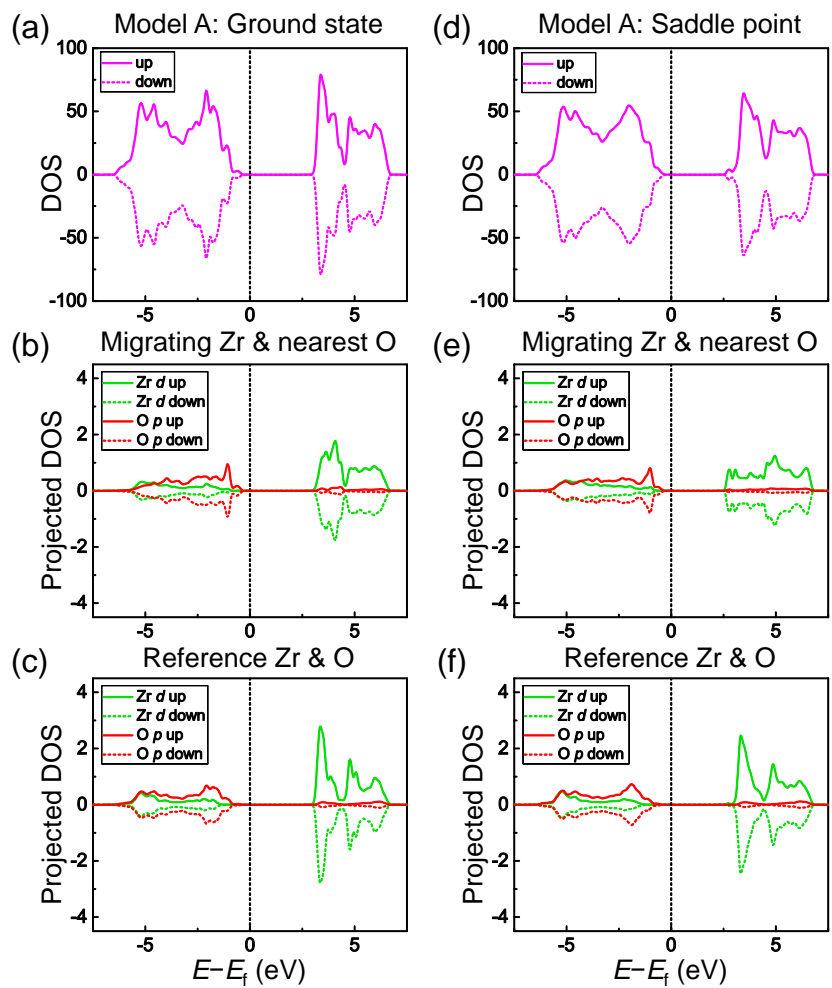

**Figure 7.2** Calculated density of states (DOS) of cubic zirconia for $Zr^{4+}$ migration **Model A**. Ground state: (a) total DOS, (b) projected DOS of (to be) migrating Zr (in green) and nearest O (in red), and (c) projected DOS of non-participating reference Zr (in green) and O (in red). Saddle-point state: (d) total DOS, (e) projected DOS of migrating Zr (in green) and nearest O (in red), and (f) projected DOS of non-participating reference Zr (in green) and O (in red). In each figure, Fermi energy is set to be zero and spin-up and spin-down states are plotted as positive and negative DOS, respectively.

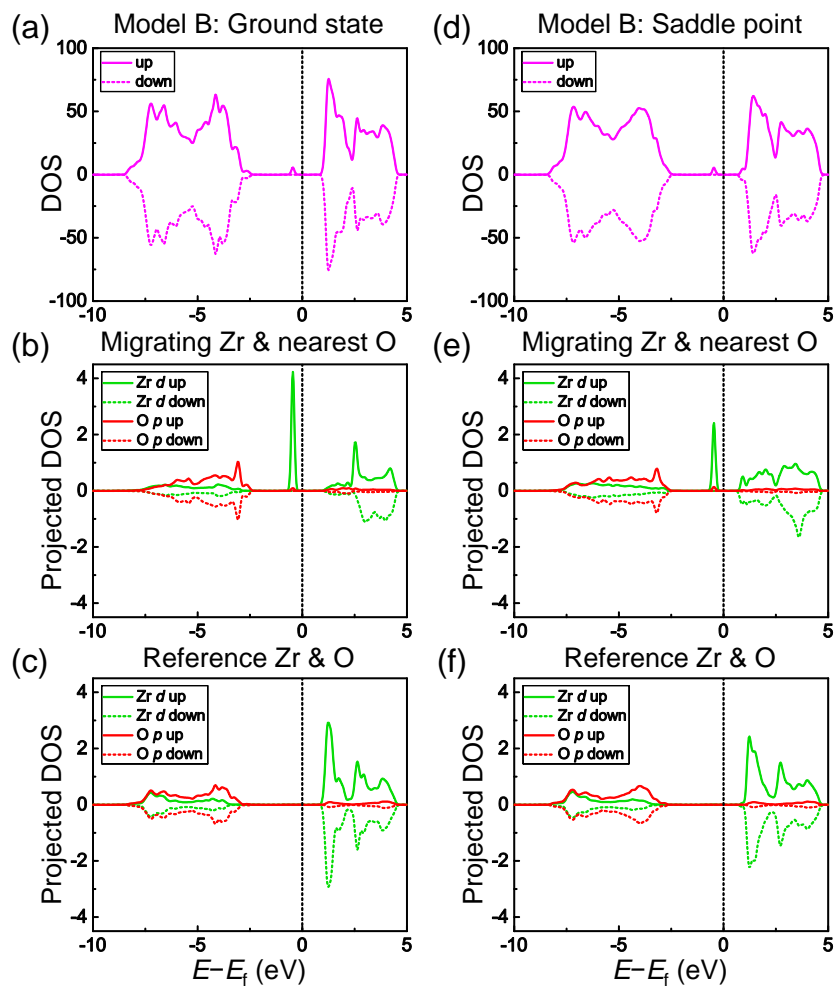

**Figure 7.3** Calculated density of states (DOS) of cubic zirconia for $Zr^{3+}$ migration **Model B**. Ground state: (a) total DOS, (b) projected DOS of (to be) migrating Zr (in green) and nearest O (in red), and (c) projected DOS of non-participating reference Zr (in green) and O (in red). Saddle-point state: (d) total DOS, (e) projected DOS of migrating Zr (in green) and nearest O (in red), and (f) projected DOS of non-participating reference Zr (in green) and O (in red). In each figure, Fermi energy is set to be zero and spin-up and spin-down states are plotted as positive and negative DOS, respectively.

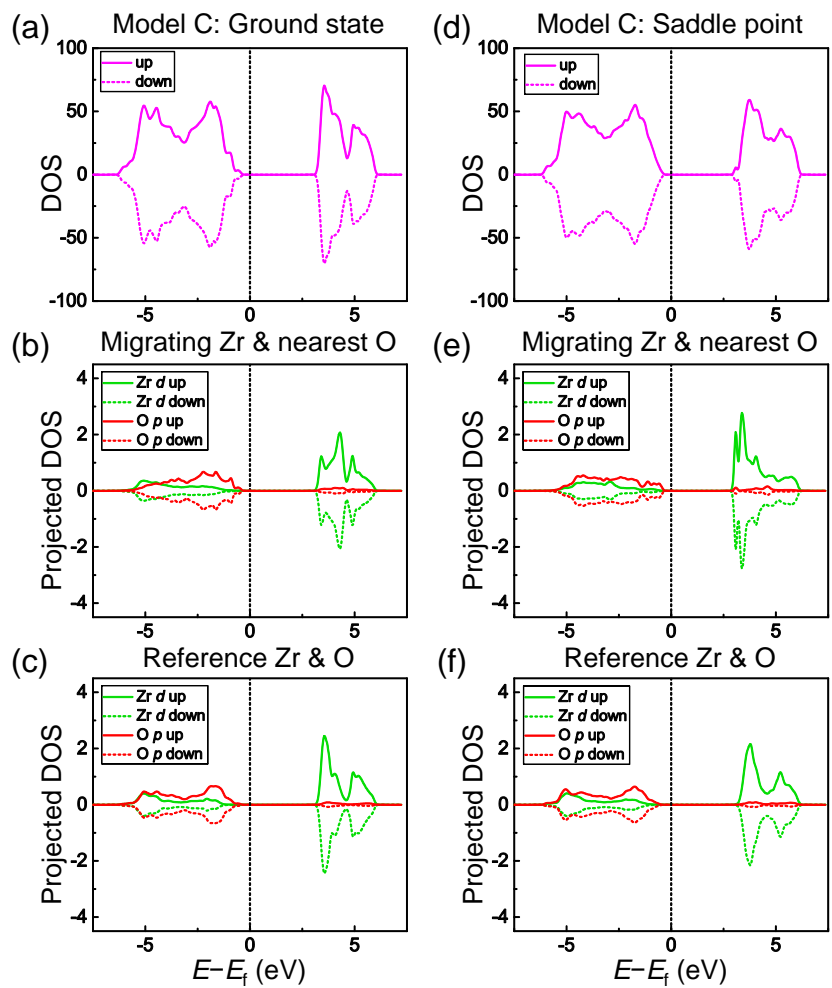

**Figure 7.4** Calculated density of states (DOS) of cubic zirconia for $Zr^{4+}$ migration **Model C**. Ground state: (a) total DOS, (b) projected DOS of (to be) migrating Zr (in green) and nearest O (in red), and (c) projected DOS of non-participating reference Zr (in green) and O (in red). Saddle-point state: (d) total DOS, (e) projected DOS of migrating Zr (in green) and nearest O (in red), and (f) projected DOS of non-participating reference Zr (in green) and O (in red). In each figure, Fermi energy is set to be zero and spin-up and spin-down states are plotted as positive and negative DOS, respectively.

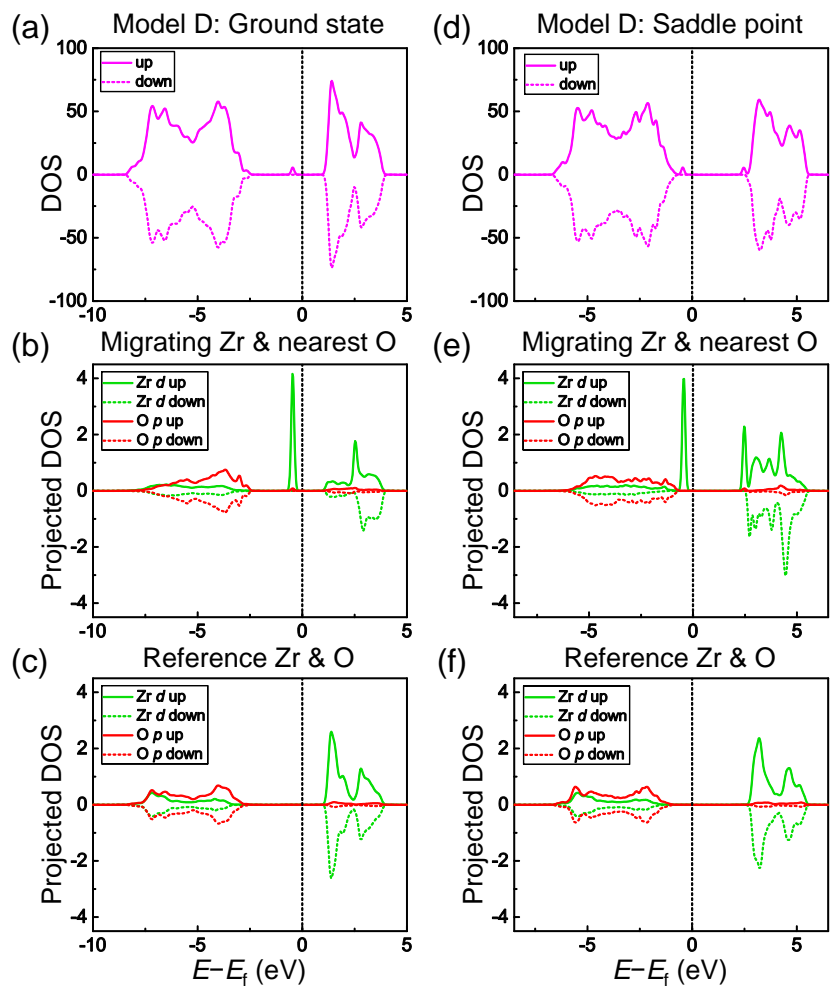

**Figure 7.5** Calculated density of states (DOS) of cubic zirconia for $Zr^{3+}$ migration **Model D**. Ground state: (a) total DOS, (b) projected DOS of (to be) migrating Zr (in green) and nearest O (in red), and (c) projected DOS of non-participating reference Zr (in green) and O (in red). Saddle-point state: (d) total DOS, (e) projected DOS of migrating Zr (in green) and nearest O (in red), and (f) projected DOS of non-participating reference Zr (in green) and O (in red). In each figure, Fermi energy is set to be zero and spin-up and spin-down states are plotted as positive and negative DOS, respectively.

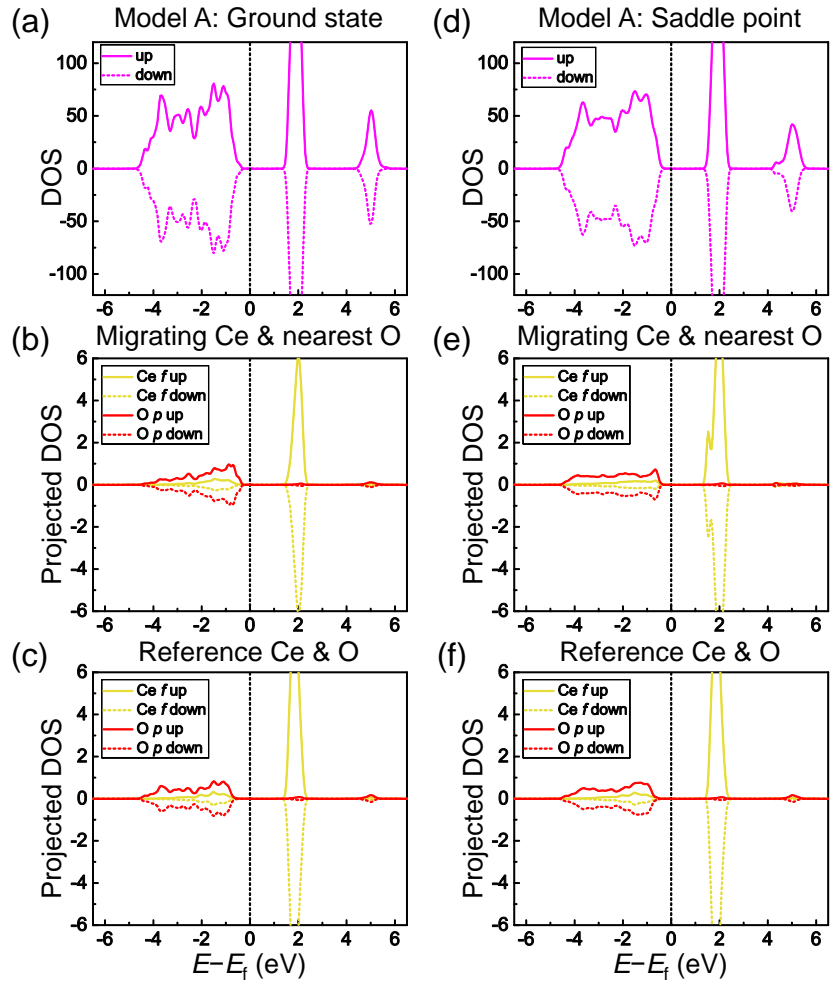

**Figure 7.6** Calculated density of states (DOS) of ceria for $Ce^{4+}$ migration **Model A**. Ground state: (a) total DOS, (b) projected DOS of (to be) migrating Ce (in yellow) and nearest O (in red), and (c) projected DOS of non-participating reference Ce (in yellow) and O (in red). Saddle-point state: (d) total DOS, (e) projected DOS of migrating Ce (in yellow) and nearest O (in red), and (f) projected DOS of non-participating reference Ce (in yellow) and O (in red). In each figure, Fermi energy is set to be zero and spin-up and spin-down states are plotted as positive and negative DOS, respectively.

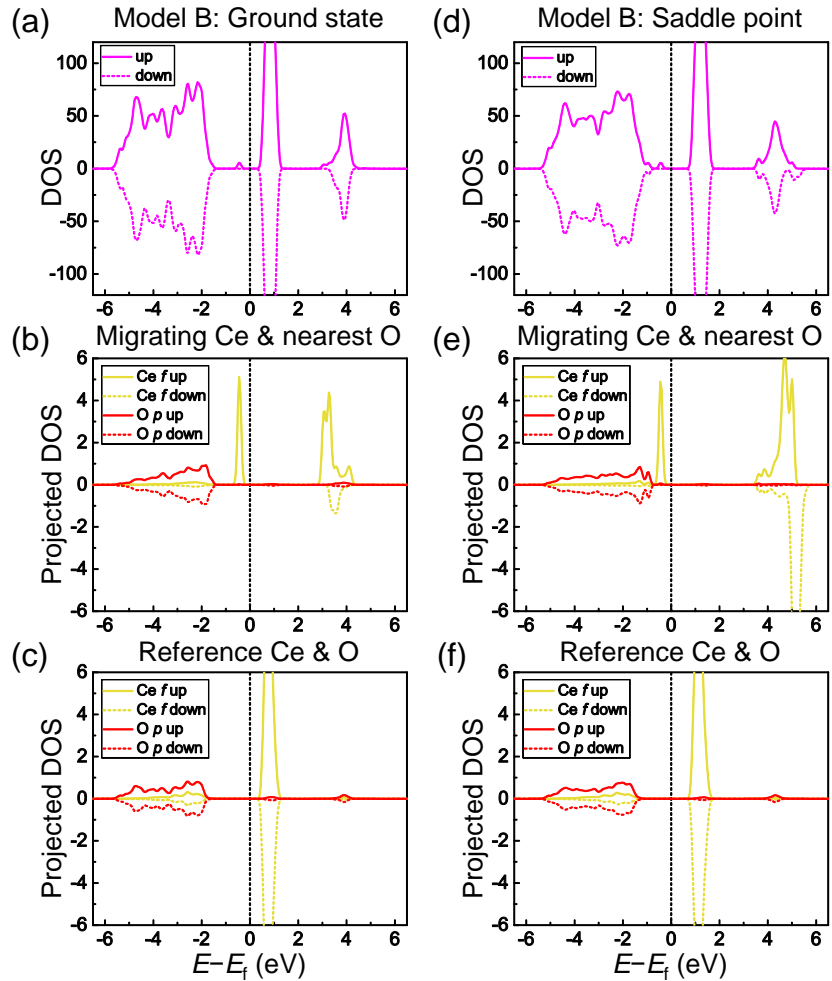

**Figure 7.7** Calculated density of states (DOS) of ceria for $Ce^{3+}$ migration **Model B**. Ground state: (a) total DOS, (b) projected DOS of (to be) migrating Ce (in yellow) and nearest O (in red), and (c) projected DOS of non-participating reference Ce (in yellow) and O (in red). Saddle-point state: (d) total DOS, (e) projected DOS of migrating Ce (in yellow) and nearest O (in red), and (f) projected DOS of non-participating reference Ce (in yellow) and O (in red). In each figure, Fermi energy is set to be zero and spin-up and spin-down states are plotted as positive and negative DOS, respectively.

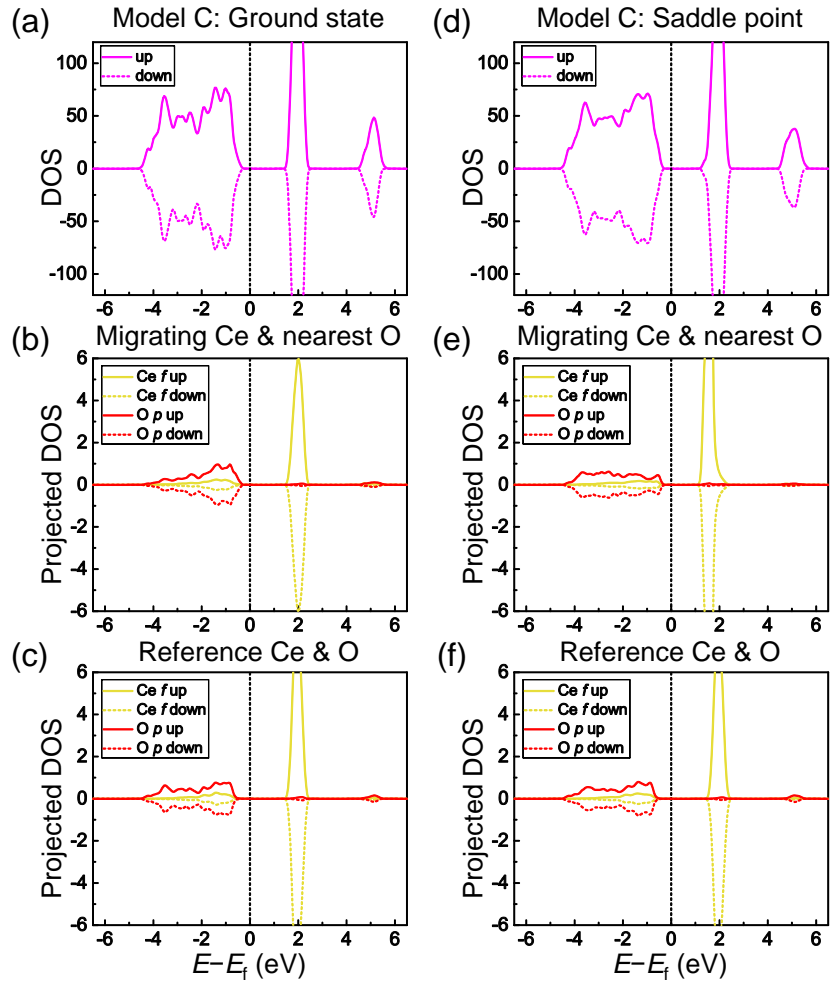

**Figure 7.8** Calculated density of states (DOS) of ceria for $Ce^{4+}$ migration **Model C**. Ground state: (a) total DOS, (b) projected DOS of (to be) migrating Ce (in yellow) and nearest O (in red), and (c) projected DOS of non-participating reference Ce (in yellow) and O (in red). Saddle-point state: (d) total DOS, (e) projected DOS of migrating Ce (in yellow) and nearest O (in red), and (f) projected DOS of non-participating reference Ce (in yellow) and O (in red). In each figure, Fermi energy is set to be zero and spin-up and spin-down states are plotted as positive and negative DOS, respectively.

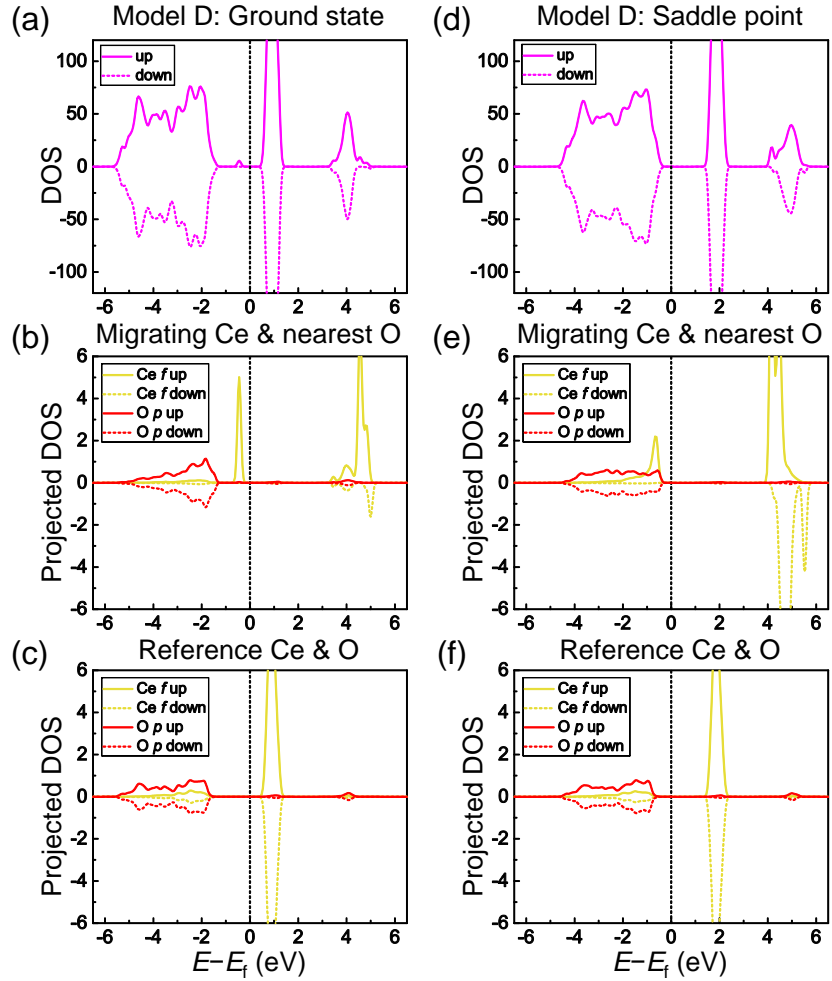

**Figure 7.9** Calculated density of states (DOS) of ceria for $Ce^{3+}$ migration **Model D**. Ground state: (a) total DOS, (b) projected DOS of (to be) migrating Ce (in yellow) and nearest O (in red), and (c) projected DOS of non-participating reference Ce (in yellow) and O (in red). Saddle-point state: (d) total DOS, (e) projected DOS of migrating Ce (in yellow) and nearest O (in red), and (f) projected DOS of non-participating reference Ce (in yellow) and O (in red). In each figure, Fermi energy is set to be zero and spin-up and spin-down states are plotted as positive and negative DOS, respectively.

## 7.4 Ti Migration in BaTiO$_3$

To compare migration of $Ti^{4+}$ and $Ti^{3+}$, with or without the aid of $V_O^{\cdot\cdot}$ or Ba vacancy $V_{Ba}''$, we follow the schemes in **Fig. 7.10a-b** (with saddle-point configurations shown in **Fig. 11**) for 100 migration and **Fig. 7.12a-b** for 110 migration. Below we present the results for 100 migration in the presence of $V_O^{\cdot\cdot}$ and 110 migration in the presence of $V_{Ba}''$. The calculated energy profiles for during migration are shown in **Fig. 7.10c** and **Fig. 7.11c**, respectively and the key data are summarized in **Table 7.2**.

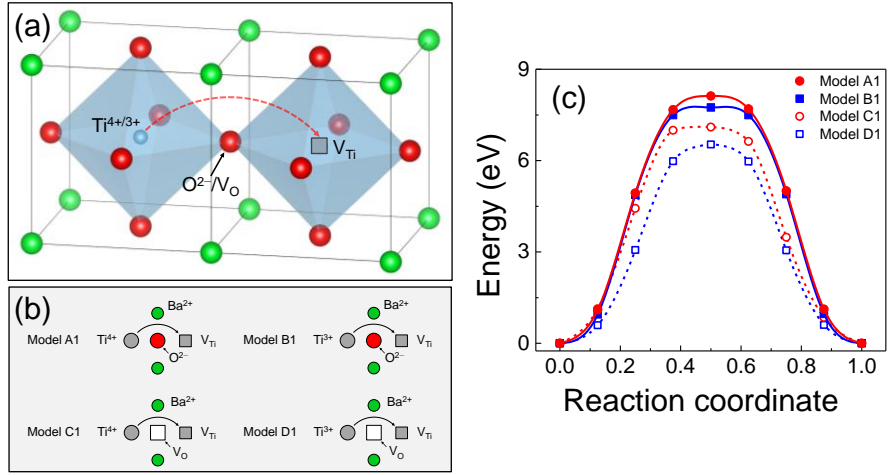

**Figure 7.10** (a) Schematics of 2 BaTiO$_3$ supercells along 100-direction with one cation vacancy $V_{Ti}$. Cation migration takes place by exchanging location of $Ti^{4+}/Ti^{3+}$ and $V_{Ti}$, with/without assistance of oxygen vacancy $V_O$. Migration found is along curved path of red dashed curve. (b) Four schematic migration models. **Model A1**: $Ti^{4+}$ exchanges with $V_{Ti}$, with no $V_O$. **Model B1**: $Ti^{3+}$ exchanges with $V_{Ti}$, with no $V_O$. **Model C1**: $Ti^{4+}$ exchanges with $V_{Ti}$, with $V_O$. **Model D1**: $Ti^{3+}$ exchanges with $V_{Ti}$, with $V_O$. (c) Cation migration energetics for **Model A1-D1** along 100-direction in BaTiO$_3$.

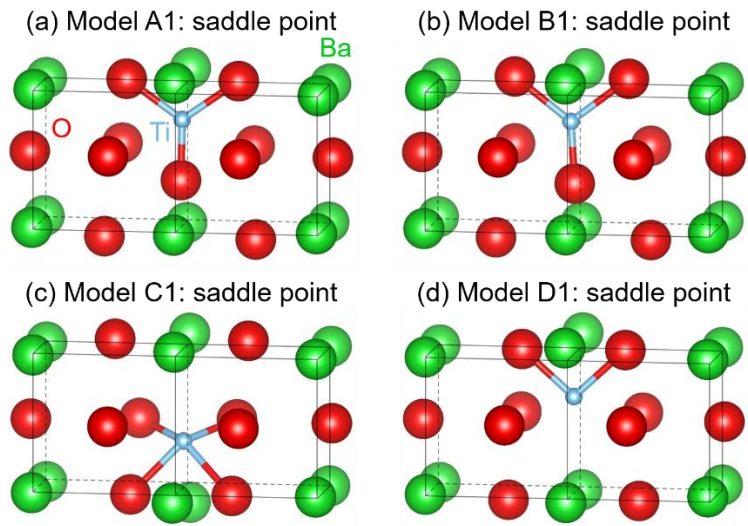

**Figure 7.11** Atomic arrangements at the saddle point in 100 migration of (a) Model A1, (b) Model B1, (c) Model C1 and (d) Model D1. Ti atom in blue, Ba atom in green and O atom in red.

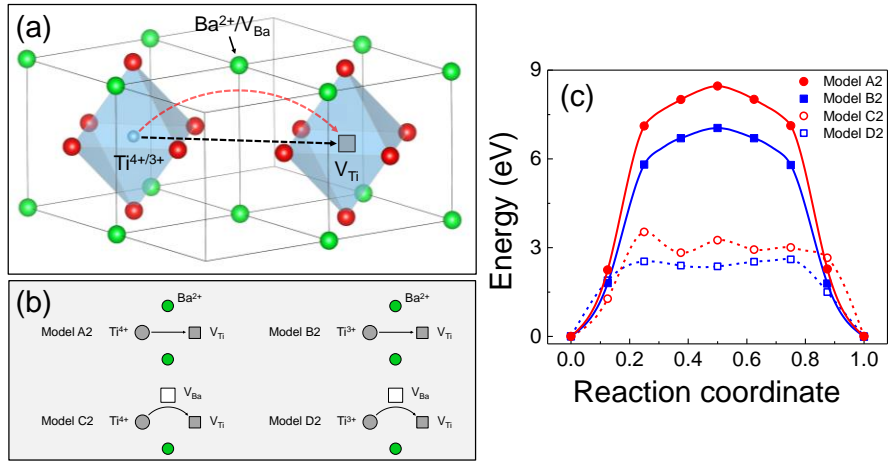

**Figure 7.12** (a) Schematics of 2 BaTiO$_3$ supercells along 110-direction with one cation vacancy $V_{Ti}$. Cation migration takes place by exchanging location of Ti$^{4+}$/Ti$^{3+}$ and $V_{Ti}$, with/without assistance of A-site vacancy $V_{Ba}$. Two migration paths found are along straight path of black dashed line and along curved path of red dashed line. (b) Four schematic migration models. **Model A2**: Ti$^{4+}$ exchanges with $V_{Ti}$, with no $V_{Ba}$. **Model B2**: Ti$^{3+}$ exchanges with $V_{Ti}$, with no $V_{Ba}$. **Model C2**: Ti$^{4+}$ exchanges with $V_{Ti}$, with $V_{Ba}$. **Model D2**: Ti$^{3+}$ exchanges with $V_{Ti}$, with $V_{Ba}$. (c) Cation migration energetics for **Model A2-D2** along 110-direction in BaTiO$_3$.

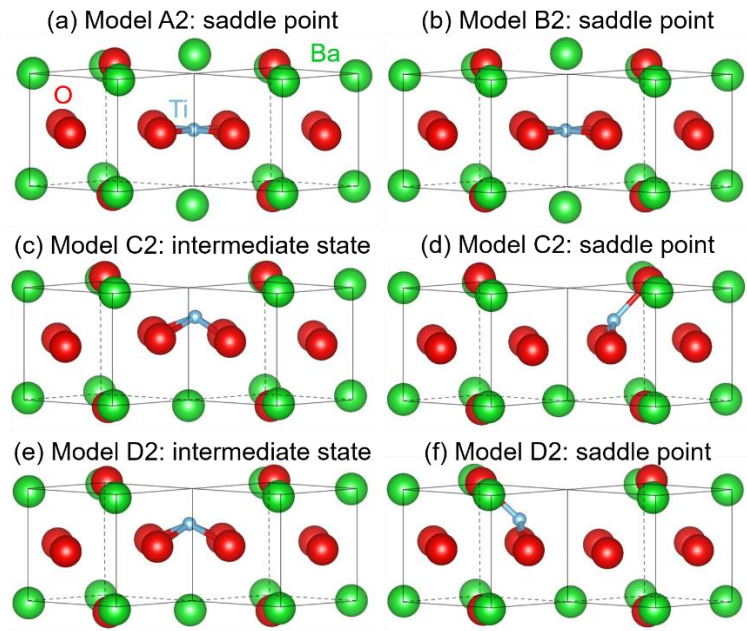

**Figure 7.13** Atomic arrangements at the saddle point of 110 migration of (a) Model A2, (b) Model B2, (d) Model C2 and (f) Model D2. Also shown are atomic arrangement at intermediate state of (c) Model C2 and (e) Model D2, which are structurally similar to saddle point of Model A2 and B2. Ti atom in blue, Ba atom in green and O atom in red.

**Table 7.2** Results of cation migration in cubic $BaTiO_3$ via a vacancy mechanism. In our DFT+$U$ calculations of defect-free cubic $BaTiO_3$, reference Ti-O bond length is 2.01 Å and Ti-Ba distance is 3.48 Å. With cation vacancy in ground state, Ti-O bond length and Ti-Ba distance in ground state, these lengths differ from reference ones. To calculate Ti's Bader charge, we exclude 8$e$ from $3s^23p^6$ inner-shell. * indicates Ti-Ba distance in intermediate state for Model C2 and D2.

| Migration direction | Model | Migration path | Migration barrier (eV) | Nearest Ti-O distance (Å) | | Nearest Ti-Ba distance (Å) | | Bader charge on migrating cation ($e$) | |
|---|---|---|---|---|---|---|---|---|---|
| | | | | Ground | Saddle | Ground | Saddle | Ground | Saddle |
| 100 | A1 | Curved | 8.12 | 1.82 | 1.68 | 3.46 | 2.77×2 | 1.84 | 1.88 |
| | B1 | Curved | 7.10 | 1.82 | 1.73 | 3.48 | 2.79×2 | 1.84 | 2.18 |
| | C1 | Curved | 7.75 | 1.81 | 1.89 | 3.43 | 2.62 | 1.80 | 1.87 |
| | D1 | Curved | 6.53 | 1.81 | 1.94 | 3.44 | 2.84/2.86 | 1.81 | 2.18 |
| 110 | A2 | Straight | 8.46 | 1.99 | 1.89 | 3.56 | 2.61×2 | 1.68 | 2.03 |
| | B2 | Straight | 7.04 | 2.04 | 1.95 | 3.50 | 2.68×2 | 2.03 | 2.19 |
| | C2 | Curved | 3.53 | 1.98 | 1.78 | 3.49 | 3.44×2 (3.10*) | 1.75 | 1.83 |
| | D2 | Curved | 2.60 | 2.06 | 1.86 | 3.48 | 3.42×2 (3.01*) | 1.96 | 2.11 |

**100 migration**

    **Model A1 ($Ti^{4+}$ migration):** In this reference case, $Ti^{4+}$ migrates along the curved path as shown in **Fig. 7.10a**, passing through the triangular window between two $Ba^{2+}$ and one $O^{2-}$. At the saddle point (**Fig. 7.11a**), the migrating Ti pushes these $Ba^{2+}$ and $O^{2-}$ away from their original locations, yet it still achieves a much shorter Ti-Ba distance (2.77 Å) and Ti-O bond length (1.68 Å) compared to the reference ones in cubic $BaTiO_3$. To screen the charge, however, the migrating Ti also pulls in two O to maintain a coordination number of three (of O) at the saddle point, which is much smaller than seen in

the ground state and in **Fig. 1a** for Zr and Ce cations. (This is also one reason why the Ti-O distance is much shorter.) There is no obvious change in the charge state (indicated by Bader charge in **Table 7.2**) or the projected DOS (**Fig. 7.14**) of the Ti during migration. The migration barrier of **Model A1** is 8.12 eV, which is unrealistically high for a compound that melts at 1900 K, probably because of the poorly screened saddle point environment for a tetravalent cation.

**Model B1 (Ti$^{3+}$ migration):** The migration path and the saddle-point configuration (**Fig. 7.11b**) of **Model B1** are essentially the same as in **Model A1**, with the major difference being the longer Ti-O bond length (see **Table 7.2**.) This is due to the larger size of Ti$^{3+}$, which is confirmed by the Bader charge (0.34$e$ increase) in **Table 7.2** and the projected DOS in **Fig. 7.15e**, which signals an impurity state has formed within the band gap at below the Fermi level. However, as also evident from the Bader charge in **Table 7.2** and the projected DOS in **Fig. 7.15b**, there is no extra electron on the target Ti at the ground state. Instead, the extra electron is delocalized causing the (Ti3$d$-dominated) CBM to fall slightly beneath the Fermi level. This reflects the high symmetry of the octahedral environment of Ti in the ground state, which makes it difficult to distort to accommodate a non-degenerate impurity state. The migration barrier of **Model B1** is 7.10 eV, about 1 eV lower than that of **Model A1**. It likely benefits from the lower valence of Ti$^{3+}$, which causes lower electrostatic energy despite poor screening in the saddle-point configuration.

**Model C1 (Ti$^{4+}$ migration aided by $V_o^{\cdot\cdot}$):** As shown in **Fig. 7.10a**, 100 migration of Ti is blocked by a lattice O. So the removal of the intervening O should allow it to happen. Nevertheless, the decrease in the migration barrier is marginal, being 7.75 eV of **Model C1** vs. 8.12 eV of **Model A1**, and the migration path is still curved along the red dashed line in **Fig. 7.10a**. This may be explained by electrostatic consideration, because a Ti cation at (½½0) finds only four Ba cations around and no O as nearest neighbors. Therefore, it is poorly screened and energetically unfavorable. As a result, the migration path veers toward the 001-direction to pull in some O, achieving a four-fold coordination. (The effective coordination number is likely to be higher judging from the longer Ti-O bond length compared to the ground state.) This is not optimal, because it brings the migrating Ti$^{4+}$ closer to Ba$^{2+}$ than in **Model A1-A2**, which increases repulsion. (The decrease of Ti-Ba distance from the ground state to the saddle-point state is 0.81 Å.) These results clearly demonstrate that having a small ionic radius and a missing oxygen neighbor cannot alleviate the high migration barrier, which undisputedly argues against the size effect.

**Model D1 ($Ti^{3+}$ migration aided by $V_O^{\cdot\cdot}$):** Once again, $Ti^{3+}$ migrates along a curved line despite the absence of the intervening O. Like in **Model B1**, the ground state does not provide electron localization (**Fig. 7.17b**) while the saddle point configuration does (**Fig. 7.17e**), which increases the Bader charge by 0.37$e$. More remarkably, the localized electron in the saddle-point state falls below the VBM, meaning electron energy is actually lower than the VBM that exists before the extra electron is added to the supercell. Clearly, the extra electron must have benefited from a much promoted Ti3$d$-O2$p$ hybridization, which is an effect not previously known for the saddle point configurations. With a lower valence than before, the migrating Ti can afford less O screening than in **Model C1**, so it only adopts two O nearest neighbors at the saddle point as shown in **Fig. 7.11d**. (The actual coordination number is likely to be higher, judging from the longer Ti-O bond length compared to that in **Model B1**, which also involves $Ti^{3+}$ with a similar Bader charge.) The migration barrier is 6.53 eV, being about 1.6 eV lower than the one in **Model A1** and is the lowest found thus far. Nevertheless, this is still an un-physically high value for $BaTiO_3$. Note that the Ti-Ba distances at the saddle point, being 2.62-2.86 Å and larger than the Ti-Ba distances in **Model A1-C1**, are still ~0.6-0.8 Å shorter than the ones in the ground state, 3.43-3.48 Å (**Table 7.2**). This suggests electrostatic repulsion from $Ba^{2+}$ is still very substantial, and it may be a limiting factor for further lowering the migration barrier. This observation motivated us to investigate 110 and the effect of Ba vacancy below.

**110 migration**

**Model A2 ($Ti^{4+}$ migration):** Here we found, for the first time for Ti, straight migration path, along the black dashed line in **Fig. 7.12a**. At the saddle point (**Fig. 7.13a**), the migrating Ti has four O neighbors in a square planar arrangement. Interestingly, Ti's Bader charge at the saddle point increases to 2.03$e$ from the ground state value, 1.68$e$, which is quite low compared to the reference case of

**Model A2** and to other cases studied here. This indicates that $Ti^{4+}$ in **Model B2** is highly ionic in the ground state, with relatively little charge sharing with neighboring O. This is also reflected in the longer Ti-O bond length and Ti-Ba distance compared to those of **Model A1**. Meanwhile, the saddle-point Ti-Ba distance (2.61 Å) suffers the largest decrease (0.95 Å) from the ground state value (3.56 Å) of all the models in **Table 2**; not surprisingly, this is correlated with the highest migration barrier of 8.46 eV in **Table 2**. So 110 migration does not offer any intrinsic advantage. Lastly, the very simple square planar Ti-O configuration allows a straightforward interpretation of the conduction band manifold in terms of crystal field splitting: While the ground state DOS of Ti in an octahedral configuration (e.g., **Fig. 7.18 c** and **f**) is split into $d_{xy}/d_{yz}/d_{zx}$ and $d_{z^2}/d_{x^2-y^2}$, the square planar configuration in the saddle-point state splits the DOS further into $d_{xz}/d_{yz}$, $d_{z^2}$, $d_{xy}$, and $d_{x^2-y^2}$ in **Fig. 7.18e.** Thus, they are expected to become relatively narrowly distributed and sometimes the highest level may be too high to be included in our plots. This picture seems to be borne out by many of the projected saddle-point Ti-DOS in **Fig. 7.15-20**.

**Model B2 ($Ti^{3+}$ migration):** Although Ti in **Model B1** is the most ionic with the smallest Bader charge seen in **Table 2**, adding an extra electron drastically changes the picture, resulting in electron localization on the target Ti in both the ground state and the saddle-point state. This is supported by the Bader charge in **Table 7.2** and the projected DOS in **Fig. 7.19b** and **e.** The state has the feature of a distinct impurity level within the band gap. Thus, $Ti^{3+}$ is very stable, which is consistent with the longer Ti-O bond lengths (**Table 7.2**) at both the ground state and the saddle-point, compared with the reference ones in **Model A2**. The migration path and saddle point configuration (**Fig. 7.13b**) of **Model B2** is essentially the same as those in **Model A2**, but with a barrier lowered by 1.42 eV, to become 7.04 eV, which may be accounted for by the down-shifting of the energy level of the impurity state. This is the same observation we had in $ZrO_2$ and $CeO_2$. But the large barrier is again unphysical for a compound that melts at 1900K, though consistent with the relatively short Ti-Ba distance (2.68 Å, decreased from the ground state by 0.82 Å), which implicates strong Coulombic repulsion.

**Model C2 ($Ti^{4+}$ migration aided by $V_{Ba}''$):** We suggested above that the large migration barrier could be due to the repulsion between Ti and Ba. This was verified here by removing a $Ba^{2+}$, forming a $V_{Ba}''$, which results in a much lower migration barrier of 3.53 eV. This is consistent with its curved path veering toward the $V_{Ba}''$ (red dashed line in **Fig. 7.12a**) and the migration energetics in **Fig. 7.12c**, which features a flat plateau where several intermediate states have very similar energies. Interestingly, the symmetric location halfway along the 110-migration path is actually an intermediate state (**Fig. 7.13c**), which has four surrounding O in a puckered square-planar configuration. (We do not rule out the possibility that this could turn out to be the saddle point in a more refined calculation, since the energy difference between the above intermediate configuration and the following saddle point configuration is quite small.) Meanwhile, the saddle point is traversed when $Ti^{4+}$ leaves the octahedral site of the ground state, entering a three-fold coordinated configuration with oxygens in **Fig. 7.13d**. This saddle point sees two nearest Ba, at 3.44 Å; the intermediate state of **Fig. 7.13c** sees only one Ba, at 3.10 Å. These distances are much longer than the ones found in other models studied thus far (**Table 7.2**). Moreover, they are much closer to the values at the ground state, which is 3.48 Å. That is, there is little change in the Ti-Ba distance during migration, a case not seen in other models explored so far. This provides further support for the strong correlation between the barrier and Ti-Ba distance-shortening during migration.

**Model D2 ($Ti^{3+}$ migration aided by $V_{Ba}''$):** Having seen a huge energetic advantage of Ti migration aided by a $V_{Ba}''$, we finally investigated whether additional cation reduction can further lower the migration barrier. As in **Model B2**, the extra electron is localized on the target Ti in both the ground state and the saddle point, as evidenced by the larger Bader charge in **Table 7.2** and the emerging impurity state in the projected DOS in **Fig. 7.21b** and **e**. As in **Model C2**, the migration path veers toward the $V_{Ba}''$, along the red dashed line schematically shown in **Fig. 7.12a**. The saddle-point and the intermediate-state configurations are almost identical to those found in **Model C2**, as shown in **Fig. 13e** and **f**. With Ti reduction, the migration barrier further decreases by more than 25% to 2.60 eV, which is the lowest in all the models studied here, and it is also accompanied by a shifting in the energy level of the impurity state. Like before, the low barrier is correlated with a very little shortening of the Ti-Ba distance during migration. Therefore, we conclude in addition to the reduction of electrostatic energy, electron localization and impurity-energy-level shifting help lowering the barrier for cation migration in BaTiO$_3$, just as in ZrO$_2$ and CeO$_2$.

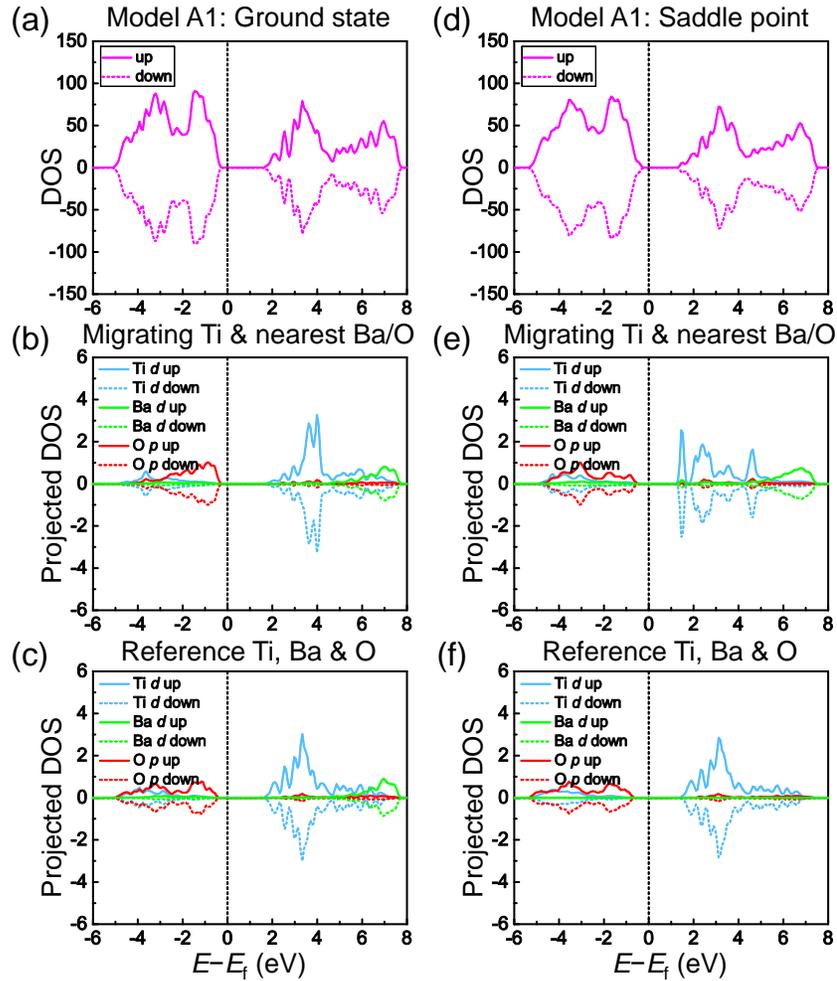

**Figure 7.14** Calculated density of states (DOS) of cubic BaTiO$_3$ for Ti$^{4+}$ 100 migration **Model A1**. Ground state: (a) total DOS, (b) projected DOS of (to be) migrating Ti (in blue), nearest Ba (in green) and O (in red), and (c) projected DOS of non-participating reference Ti (in blue), Ba (in green) and O (in red). Saddle-point state: (d) total DOS, (e) projected DOS of migrating Ti (in blue), nearest Ba (in green) and O (in red), and (f) projected DOS of non-participating reference Ti (in blue), Ba (in green) and O (in red) at the saddle-point configuration. In each figure, Fermi energy is set to be zero and spin-up and spin-down states are plotted as positive and negative, respectively.

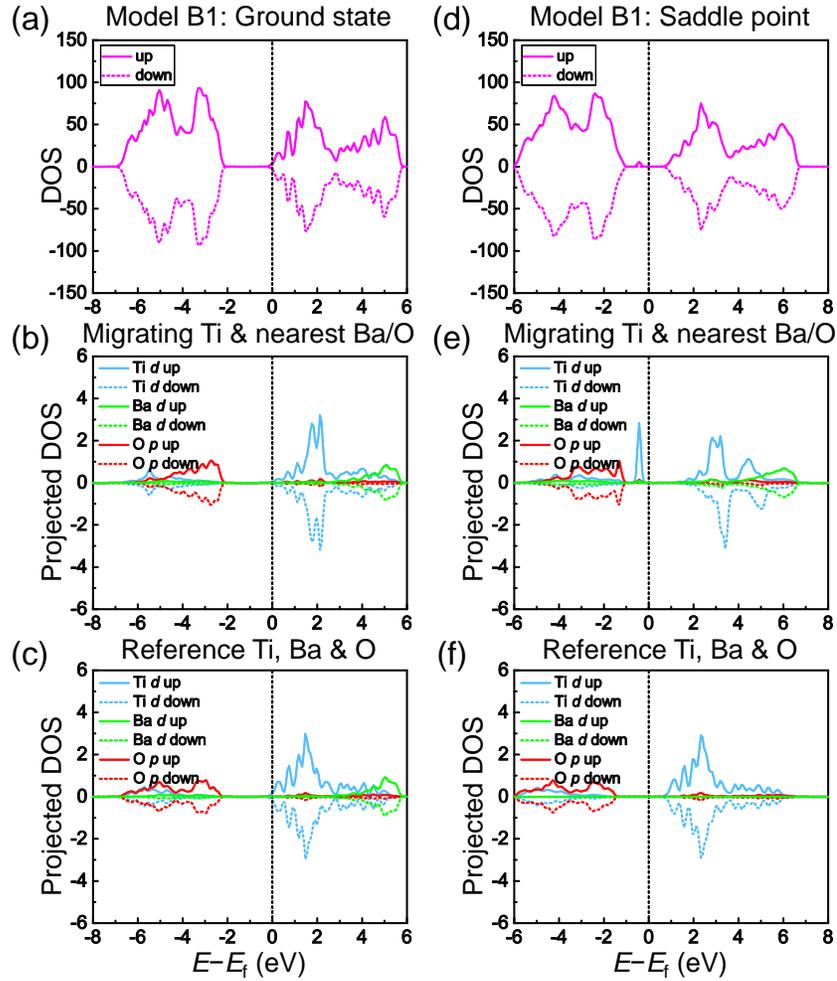

**Figure 7.15** Calculated density of states (DOS) of cubic BaTiO$_3$ for Ti$^{3+}$ 100 migration **Model B1**. Ground state: (a) total DOS, (b) projected DOS of (to be) migrating Ti (in blue), nearest Ba (in green) and O (in red), and (c) projected DOS of non-participating reference Ti (in blue), Ba (in green) and O (in red). Saddle-point state: (d) total DOS, (e) projected DOS of migrating Ti (in blue), nearest Ba (in green) and O (in red), and (f) projected DOS of non-participating reference Ti (in blue), Ba (in green) and O (in red) at the saddle-point configuration. In each figure, Fermi energy is set to be zero and spin-up and spin-down states are plotted as positive and negative, respectively.

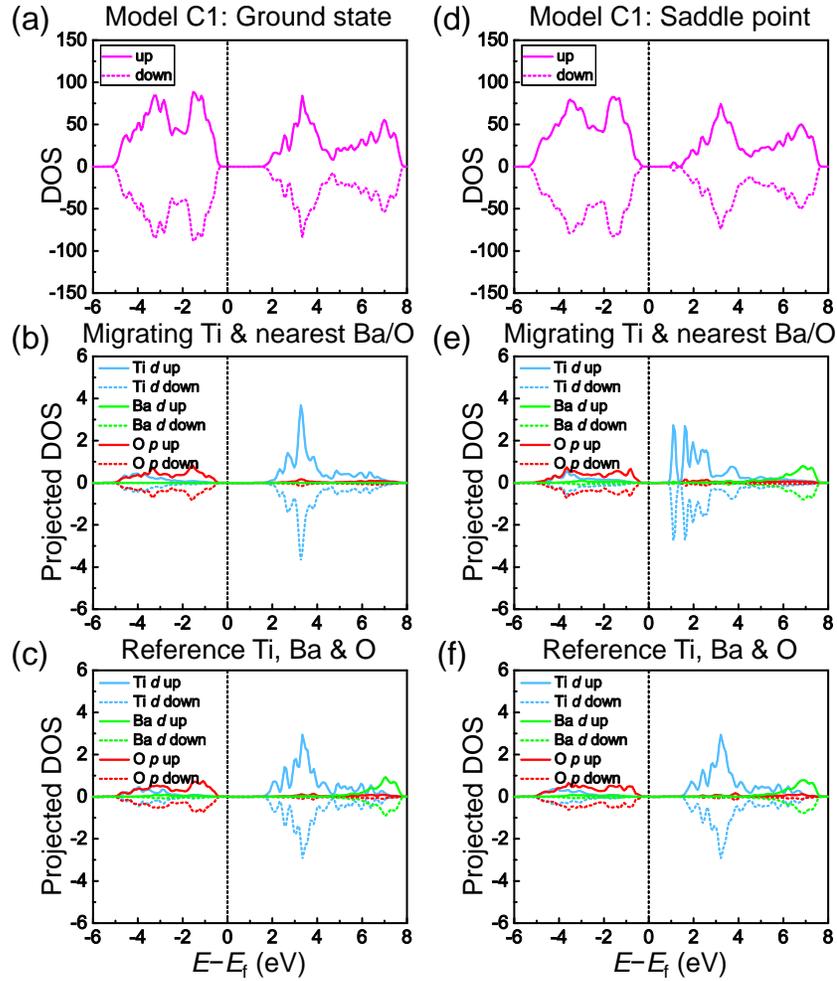

**Figure 7.16** Calculated density of states (DOS) of cubic BaTiO$_3$ for Ti$^{4+}$ 100 migration **Model C1**. Ground state: (a) total DOS, (b) projected DOS of (to be) migrating Ti (in blue), nearest Ba (in green) and O (in red), and (c) projected DOS of non-participating reference Ti (in blue), Ba (in green) and O (in red). Saddle-point state: (d) total DOS, (e) projected DOS of migrating Ti (in blue), nearest Ba (in green) and O (in red), and (f) projected DOS of non-participating reference Ti (in blue), Ba (in green) and O (in red) at the saddle-point configuration. In each figure, Fermi energy is set to be zero and spin-up and spin-down states are plotted as positive and negative, respectively.

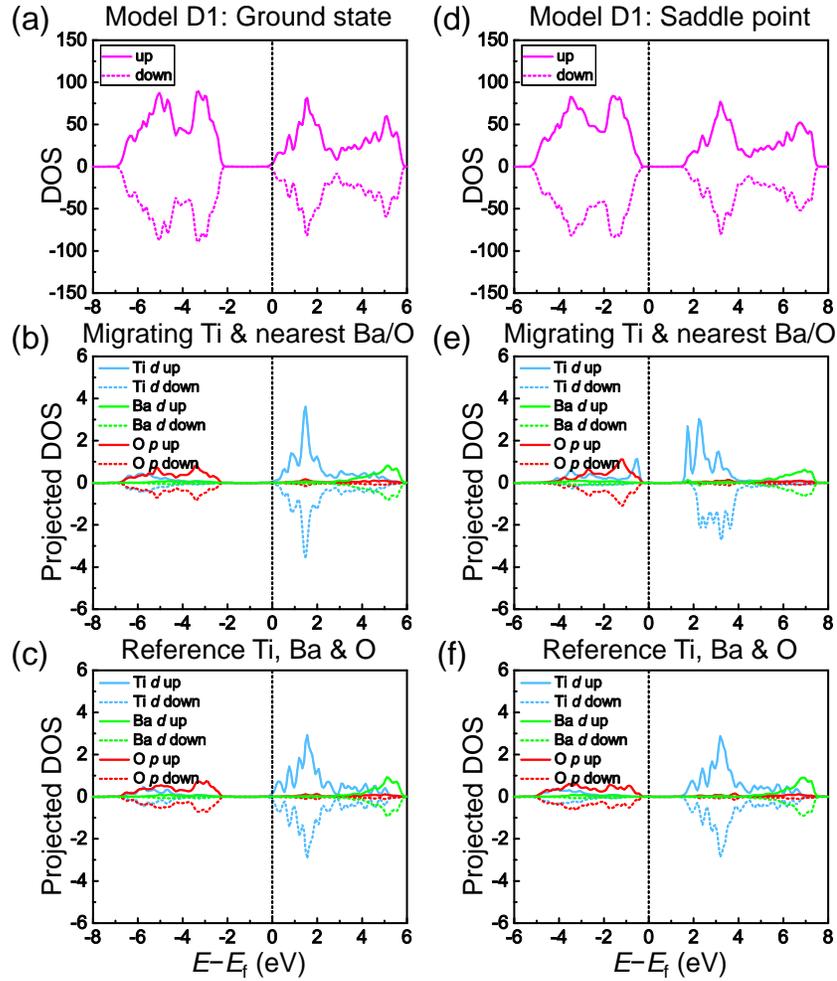

**Figure 7.17** Calculated density of states (DOS) of cubic $BaTiO_3$ for $Ti^{3+}$ 100 migration **Model D1**. Ground state: (a) total DOS, (b) projected DOS of (to be) migrating Ti (in blue), nearest Ba (in green) and O (in red), and (c) projected DOS of non-participating reference Ti (in blue), Ba (in green) and O (in red). Saddle-point state: (d) total DOS, (e) projected DOS of migrating Ti (in blue), nearest Ba (in green) and O (in red), and (f) projected DOS of non-participating reference Ti (in blue), Ba (in green) and O (in red) at the saddle-point configuration. In each figure, Fermi energy is set to be zero and spin-up and spin-down states are plotted as positive and negative, respectively.

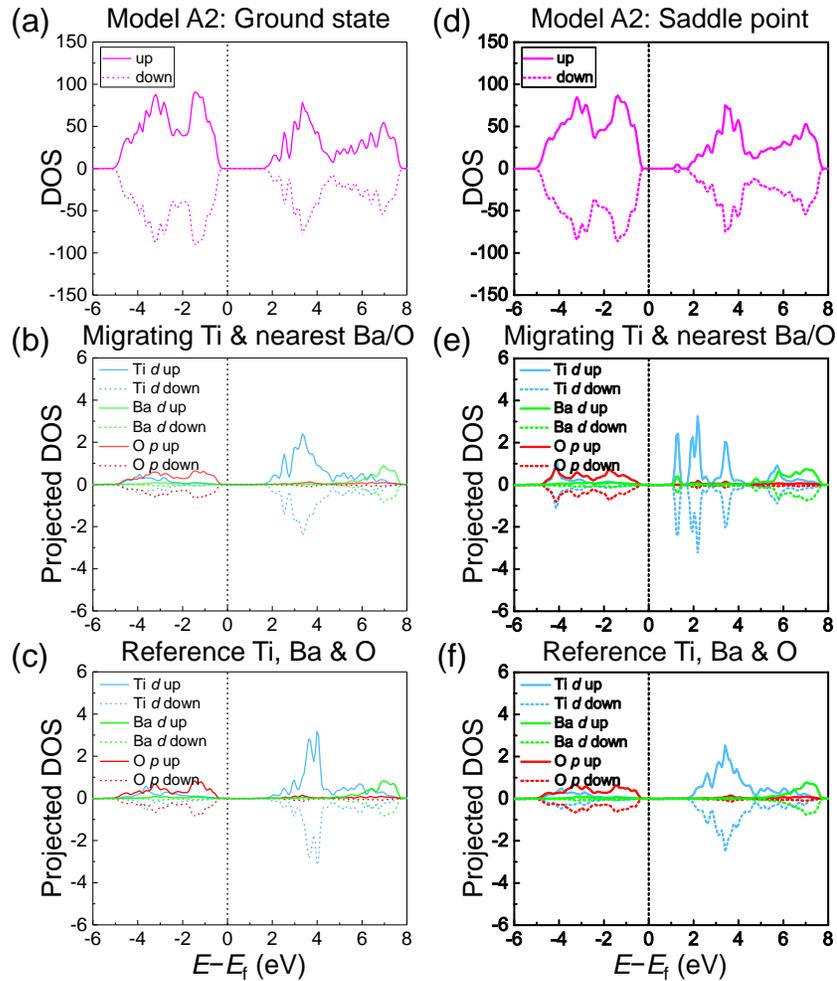

**Figure 7.18** Calculated density of states (DOS) of cubic BaTiO$_3$ for Ti$^{4+}$ 110 migration **Model A2**. Ground state: (a) total DOS, (b) projected DOS of (to be) migrating Ti (in blue), nearest Ba (in green) and O (in red), and (c) projected DOS of non-participating reference Ti (in blue), Ba (in green) and O (in red). Saddle-point state: (d) total DOS, (e) projected DOS of migrating Ti (in blue), nearest Ba (in green) and O (in red), and (f) projected DOS of non-participating reference Ti (in blue), Ba (in green) and O (in red) at the saddle-point configuration. In each figure, Fermi energy is set to be zero and spin-up and spin-down states are plotted as positive and negative, respectively.

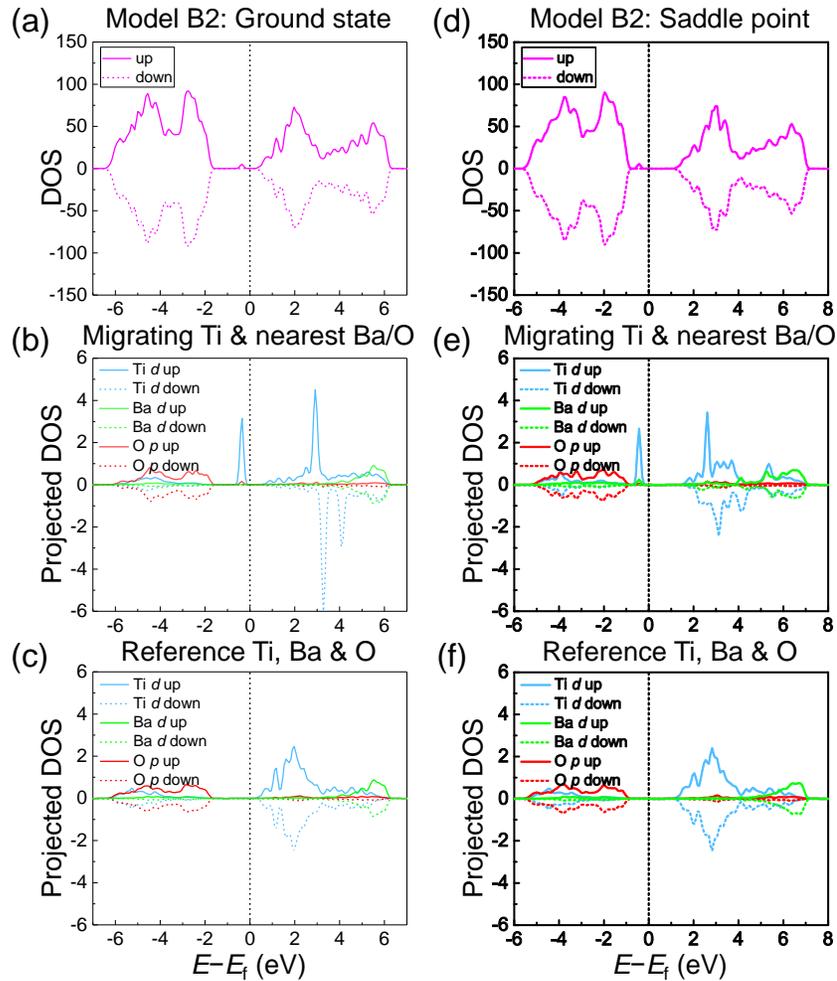

**Figure 7.19** Calculated density of states (DOS) of cubic BaTiO$_3$ for Ti$^{3+}$ 110 migration **Model B2**. Ground state: (a) total DOS, (b) projected DOS of (to be) migrating Ti (in blue), nearest Ba (in green) and O (in red), and (c) projected DOS of non-participating reference Ti (in blue), Ba (in green) and O (in red). Saddle-point state: (d) total DOS, (e) projected DOS of migrating Ti (in blue), nearest Ba (in green) and O (in red), and (f) projected DOS of non-participating reference Ti (in blue), Ba (in green) and O (in red) at the saddle-point configuration. In each figure, Fermi energy is set to be zero and spin-up and spin-down states are plotted as positive and negative, respectively.

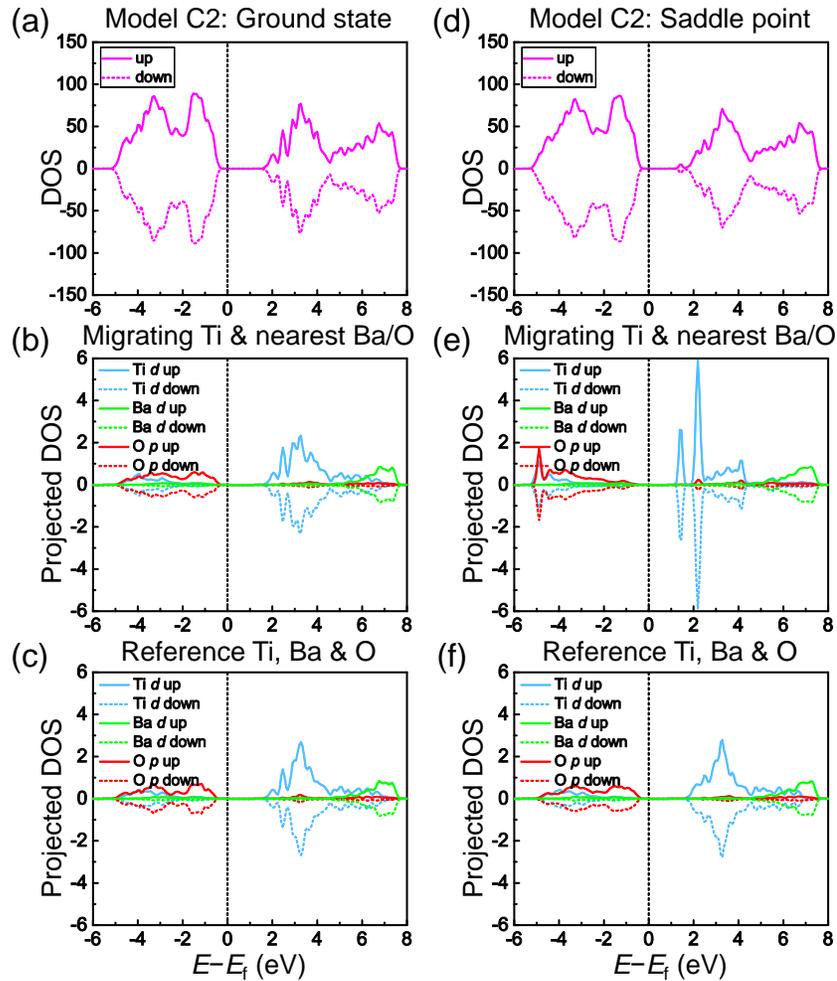

**Figure 7.20** Calculated density of states (DOS) of cubic BaTiO$_3$ for Ti$^{4+}$ 110 migration **Model C2**. Ground state: (a) total DOS, (b) projected DOS of (to be) migrating Ti (in blue), nearest Ba (in green) and O (in red), and (c) projected DOS of non-participating reference Ti (in blue), Ba (in green) and O (in red). Saddle-point state: (d) total DOS, (e) projected DOS of migrating Ti (in blue), nearest Ba (in green) and O (in red), and (f) projected DOS of non-participating reference Ti (in blue), Ba (in green) and O (in red) at the saddle-point configuration. In each figure, Fermi energy is set to be zero and spin-up and spin-down states are plotted as positive and negative, respectively.

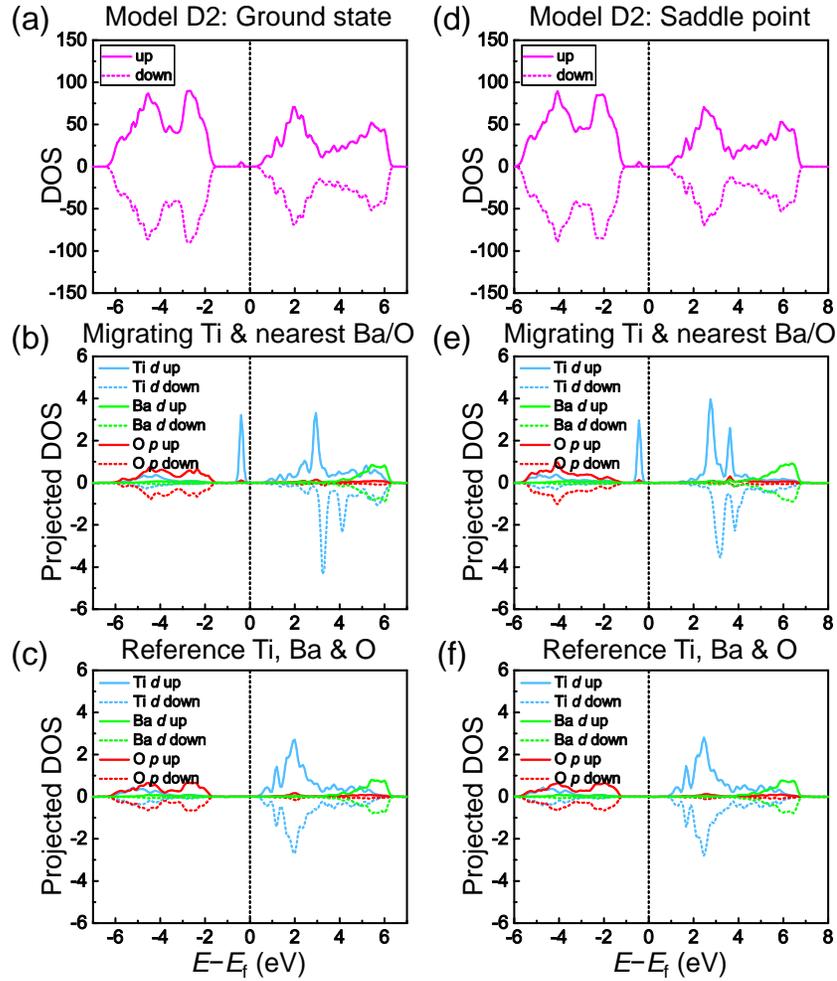

**Figure 7.21** Calculated density of states (DOS) of cubic BaTiO$_3$ for Ti$^{3+}$ 110 migration **Model D2**. Ground state: (a) total DOS, (b) projected DOS of (to be) migrating Ti (in blue), nearest Ba (in green) and O (in red), and (c) projected DOS of non-participating reference Ti (in blue), Ba (in green) and O (in red). Saddle-point state: (d) total DOS, (e) projected DOS of migrating Ti (in blue), nearest Ba (in green) and O (in red), and (f) projected DOS of non-participating reference Ti (in blue), Ba (in green) and O (in red) at the saddle-point configuration. In each figure, Fermi energy is set to be zero and spin-up and spin-down states are plotted as positive and negative, respectively.

## 7.5 Discussion

### 7.5.1 Reduction enhanced cation diffusion

Our study revealed that the lattice ion vacancies and reduction can significantly lower the migration barrier of cations. Since the lattice vacancies assumed in our study, $V_O^{\bullet\bullet}$ and $V_{Ba}''$, form and diffuse easily in these compounds, they are likely to be available around a $M^{4+}$ vacancy. Therefore, the reduction effect that is known to enhance cation diffusion in fluorite structure oxides and perovskites can now be understood. This reduction effect is rooted in an energetic argument: There is a substantial lowering of the energy of the saddle-point state, by about 1 eV due to reduction alone, and much more when the synergistic effect of lattice vacancy is included, especially if it is a cation vacancy on another cation sublattice, like Ba on the A-site sublattice. Therefore, this energetic mechanism is much more powerful than the defect chemistry argument that is based on the law of mass action, which is entropic in nature.

More specifically, cation diffusion in reduced zirconia, ceria and $BaTiO_3$ is likely to proceed as follows: with a very small hopping barrier of both $V_O^{\bullet\bullet}$ and electrons of about 0.5 eV or lower, there should be enough time for $V_O^{\bullet\bullet}$ and electron to come to a cation vacancy and optimize their configurations around a surrounding cation before the latter makes a successful exchange (with a ~3 eV migration barrier) to the adjacent cation vacancy. Since the concentration of cation vacancies is very low, the above picture holds even with a dilute concentration of electrons and O/Ba vacancies. The above mechanism is directly supported by the following experimental observations. (a) In yttria stabilized zirconia, which has a wide band gap of about 5 eV and is difficult to reduce, the grain boundary mobility can be increased by ~2 times by $H_2$ reduction and >1,000 times by a severe electrical reduction[2]. (b) In undoped ceria (ceria can be easily reduced due to the presence of a stable +3 valence state), the grain boundary mobility in air is ~2 times faster than that in pure oxygen[1]; in Gd-doped ceria, the grain boundary mobility is ~400 times faster than that in air[2]. Enhanced sintering[4,5], creep[3], viscoelastic property[6], and cation inter-diffusion[7] have also been reported in ceria under a reducing condition. (c) In $BaTiO_3$, although A-site vacancies may be difficult to form by thermal activation alone, oxygen vacancies and electrons are commonplace, and A-site vacancies are relatively easily induced by donor doping, such as A-site La doping. These doping schemes are known to greatly facilitate Ti diffusion in $SrTiO_3$.[16,17] Presumably, Ti diffusion will be even faster in such materials in a reducing atmosphere.

7.5.2   Ionics and beyond

Our finding that the charge effect is of paramount importance is within the realm of classical theory of ionic compound, which stresses the key influence of Madelung energy. Thus, cation reduction lowers the migration barrier, and preventing cations from approaching each other— keeping the Ti-Ba distance unchanged during migration by creating Ba vacancy nearby—is especially beneficial. (The latter case can also be formally viewed as allowing Ti to migrate in an entirely screened pathway, starting from the $TiO_6$ ground state, and passing through the $TiO_{12}$ saddle point located at the vacant Ba site, and ending at another $TiO_6$ ground state.) In contrast, the size effect as exemplified by the oxygen-vacancy effect is relatively small (0.33 eV for $Zr^{4+}$, 0.36 eV for $Ce^{4+}$, and 0.37 eV for $Ti^{4+}$ in 100 migration, which is a factor 4 to 10 smaller than the charge effects), probably because tetravalent cations are already relatively small and thus relatively easy to pass through the saddle point if it were not for the electrostatic repulsion. The fact that curved migration path is favored by Ti 100 migration even when an intervening O is replaced by a $V_O^{\cdot\cdot}$ is an excellent testimony that the size effect is rather insignificant in the face of the electrostatic repulsion. However, as shown by the larger size effect for $M^{3+}$ (0.73 eV for $Zr^{3+}$, 1.71 eV for $Ce^{3+}$, and 0.57 eV for $Ti^{3+}$ in 100 migration) than for $M^{4+}$, and by the many examples in the projected DOS of the saddle-point state illustrating an unmistaken tendency for the extra electron to localize on the migrating cation in the saddle-point configuration, there is also an electronic aspect beyond the electrostatic and size considerations. This will undoubtedly lower the migration energy, and being quantum mechanical in nature it lies beyond the classical theory of ionic compounds.

This effect is quite general: The energy level of the localized state is a $4d$ state for Zr, a $4f$ state for Ce, and a $3d$ state for Ti. One of the reasons for localization is that the saddle point has a lower symmetry and fewer surrounding anions than the ground state. Therefore, the cation orbitals are further split into finer, sharper levels, making it easier to localize once hybridization with oxygen brings them below the Fermi level. This is very clear in the case of $BaTiO_3$, where we have identified a transition from the octahedral environment in the ground state to a square-planar, puckered square-planar, or other relatively similar low-symmetry, low-coordination configurations (e.g., a square pyramid) in the saddle point. Having identified the symmetry of the energy levels, one can immediately estimate the energy lowering by comparing the localized level and the corresponding delocalized level remaining in the conduction band manifold, noting that they have the same symmetry, comprise of the same orbital, but have opposite spins. Comparing the saddle-point DOS of $Ti^{4+}$ and $Ti^{3+}$ in this way, we estimate the lowering is about 2 eV, or slightly less than the CBM-VBM gap. In $ZrO_2$ and $CeO_2$, the extra electron is already localized in the ground state, but a very large lowering of the saddle-point projected DOS is also seen, about 1.5 eV in $Zr^{3+}$ and 1.2 eV in $Ce^{3+}$. There is no question that reduction will lower the migration barrier by this effect, which is bonding in nature instead of electrostatic in nature.

7.5.3 Negative $U$ center[36]

To understand the energetic effect and especially the synergistic effect, one need to examine more closely the energy of the localized electron, from reduction, at the saddle point. Generally, the expectation is that the extra electron by occupying a higher-energy, hitherto-unoccupied DOS must increase the total energy. That is, there is a positive $U$, which is referred to as the Hubbard $U$, and it may be attributed to the on-site Coulomb energy. On the other hand, if adding an electron actually causes the hitherto-unoccupied level to drop below the Fermi level, and indeed to fall below the highest energy of the hitherto occupied states, then one may regard it a case of negative $U$, and the site that allows this to happen is a negative-$U$ center. In our calculations, the highest energy of the hitherto occupied state is the VBM, essentially entirely of O2$p$ nature. Therefore, if the localized state (the impurity state in the saddle-point state) drops into the VBM after the addition of one extra electron, then the reduced cation situated at the saddle point must be a negative-$U$ center. Using this method, we have identified at least two negative-$U$ centers in our calculations: $Ce^{3+/4+}$ in **Model D**, and $Ti^{3+/4+}$ in **Model D1**, both having an oxygen vacancy next to the saddle point. In other cases, the most positive $U$ (2.5 eV) for $Zr^{3+/4+}$ is seen in **model B**, whereas it is just slightly positive (0.1-0.2 eV) for $Zr^{3+/4+}$ in

**Model D** and Ce$^{3+/4+}$ in **Model B**; in BaTiO$_3$, it is all slightly positive. But even though they fail to qualify as negative-$U$ centers, they still feature a $U$ less than the band gap, which exceeds 3 eV in all cases. Therefore, there is still some unaccounted-for energy-lowering when adding an electron to the saddle-point cation.

The above results may be understood as follows. The saddle-point configuration is elastically soft and fluid; indeed, by definition, it has a negative elastic modulus in the migration direction. So the addition of an extra electron, which may require atomic displacement nearby to achieve hybridization with O2$p$ orbitals, can achieve such displacement rather easily. Since lattice relaxation lowers the energy, and such relaxation is predicated in this case by the introduction of an additional electron, the situation amounts to a negative electron-phonon interaction, which if so large as to exceed the on-site Coulomb energy will result in a negative $U$. This is apparently the case in all three compounds when a neighboring lattice site is vacant (**Model D** and **Model D1**). In this way, the synergistic effect of lattice vacancy and reduction is now better understood. Lastly, we also see that a cation (Ba) vacancy despite its huge effect on lowering the migration barrier does not lead to a negative $U$. This is understandable in our picture: As mentioned previously, the saddle point in this case may be regarded as very well surrounded by oxygens, so its environment is relatively stiff and does not lend much to facilitating electron-phonon interaction.

## 7.6 Conclusions

(1) The lowest cation migration barriers to a cation vacancy according to first-principles calculations are 3.17 eV in cubic ZrO$_2$ for Zr$^{3+}$ aided by oxygen vacancy, 3.28 eV in CeO$_2$ for Ce$^{3+}$ aided by oxygen vacancy, and 2.60 eV in BaTiO$_3$ for Ti$^{3+}$ aided by A-site vacancy. These results are relevant since these materials are known for having more than enough electrons, oxygen vacancies, and A-site vacancies, which can easily migrate to the vicinity of M$^{4+}$ vacancy.

(2) The effects of cation reduction and lattice vacancy is mostly due to electrostatic consideration. The size effect due to oxygen vacancy alone is quite small, but it is considerably amplified when the cation is reduced, which creates a negative-$U$ center at the saddle point. More broadly, a strong ubiquitous electronic influence is seen in the strong tendency to localize the extra electron on the saddle-point to take advantage of the orbital level splitting, better hybridization with O2$p$, and the facility of a soft local environment that permits lattice distortion to maximize hybridization.

(3) A-site vacancy in BaTiO$_3$ is a powerful promoter for Ti migration, because it offers a saddle point fully screened by anion that has rather low migration barrier. This effect may be generalized: Cation vacancies in another interpenetration sublattice will enhance cation diffusion.